\documentclass{article}
\usepackage{spconf}

		\usepackage[utf8]{inputenc}
		\usepackage{cite}
		\usepackage{graphicx}
		\usepackage{tikz}
		\usepackage{pgfplots}
		\usepgfplotslibrary{groupplots}
			\pgfplotsset{compat=newest,legend style = {font=\footnotesize}}
		\usepackage{amsmath,amssymb,amsthm}
		\usepackage{url}
		\usepackage{etoolbox}
		\usepackage{xpatch}
		\usepackage{algorithmic}
			\makeatletter
			\xpatchcmd{\algorithmic}{\itemsep\z@}{\itemsep=8pt plus4pt}{}{}
			\makeatother
		\usepackage{array}
		\usepackage[caption=false,font=normalsize,labelfont=sf,textfont=sf]{subfig}
		\usepackage{hyperref}

\def\reg{\mathcal{R}}
\def\reals{\mathbb{R}}
\def\prox{\operatorname{\mathrm{prox}}}

\newcommand{\listint}[1]{\left\lbrace 1, 2, \dots, #1 \right\rbrace}

\title{SpotNet -- Learned iterations for cell detection in image-based immunoassays}
\name{ Pol del Aguila Pla$\dagger$,Vidit Saxena$\dagger$\thanks{$\dagger$ These two authors contributed equally to the paper and share the first authorship.}, and Joakim Jaldén\thanks{Vidit Saxena was partially supported by the Wallenberg AI, Autonomous Systems and Software Program (WASP) funded by the Knut and Alice Wallenberg Foundation. The authors would like to thank Dr. C. Smedman for providing the expert labeling of synthetic data.}}
\address{ Department of Information Science and Engineering\\
		 School of Electrical Engineering and Computer Science\\
		 KTH Royal Institute of Technology,  Stockholm 11428, Sweden \\
		 \url{[poldap,vidits,jalden]@kth.se}}

		\def\figs{figs}
				
\begin{document}
\ninept

\maketitle

\begin{abstract}
	Accurate cell detection and counting in the image-based ELISpot and FluoroSpot immunoassays is a challenging task. Recently proposed methodology matches human accuracy by leveraging knowledge of the underlying physical process of these assays and using proximal optimization methods to solve an inverse problem. Nonetheless, thousands of computationally expensive iterations are often needed to reach a near-optimal solution. In this paper, we exploit the structure of the iterations to design a parameterized computation graph, \emph{SpotNet}, that learns the patterns embedded within several training images and their respective cell information. Further, we compare SpotNet to a convolutional neural network layout customized for cell detection. We show empirical evidence that, while both designs obtain a detection performance on synthetic data far beyond that of a human expert, SpotNet is easier to train and obtains better estimates of particle secretion for each cell.
\end{abstract}


\section{Introduction}
\label{sec:intro}

\begin{figure}
  \centering
  \includegraphics[scale=1.3235]{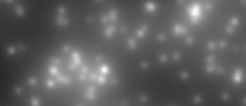}

    \vspace{-5pt}

  \caption{ \label{fig:example_image} \small Synthetically generated image of a FluoroSpot assay, illustrating the common challenges for cell detection in such images, i.e., overlapping spots of heterogeneous size and shape. Generated using the observation model derived in \cite{AguilaPla2017} and following the procedure described in Sec.~\ref{sec:exp}. }
\end{figure}

Image-based immunoassays, such as the industry-standard ELISpot and its multiplex version FluoroSpot, generate images with heterogeneous and overlapping spots centered at the location of particle-secreting cells, e.g, see Fig.~\ref{fig:example_image}. Immunoassay image analysis aims to count and localize the cells that appear in a measured image, as well as to estimate the particle secretion profile over time for each cell. In~\cite{AguilaPla2017,AguilaPla2017a,AguilaPla2018}, our group derived a mathematical framework that codifies the particle reaction-diffusion-adsorption-desorption process governing image formation in these immunoassays, and developed a novel technique to estimate the cell information from measured images. This approach matches the detection performance of a human expert, and has inspired a recent commercial product for automated cell detection and secretion profile estimation \cite{IRIS2018}. Despite its success, the approach is iterative, and typically requires thousands of computationally complex iterations to achieve near-optimal results. In this paper, we propose a learning-based alternative that relies on offline training to significantly reduce the computational complexity. Further, we provide empirical evidence that the proposed methodology provides highly accurate results on synthetic data, outperforming both a human expert and convolutional neural networks (CNN) that are state-of-the-art in image pattern recognition.

\begin{figure*}
	\centering 
		\includegraphics[scale=.435,keepaspectratio=true,clip,trim=20pt 7.3in 20pt 15pt]{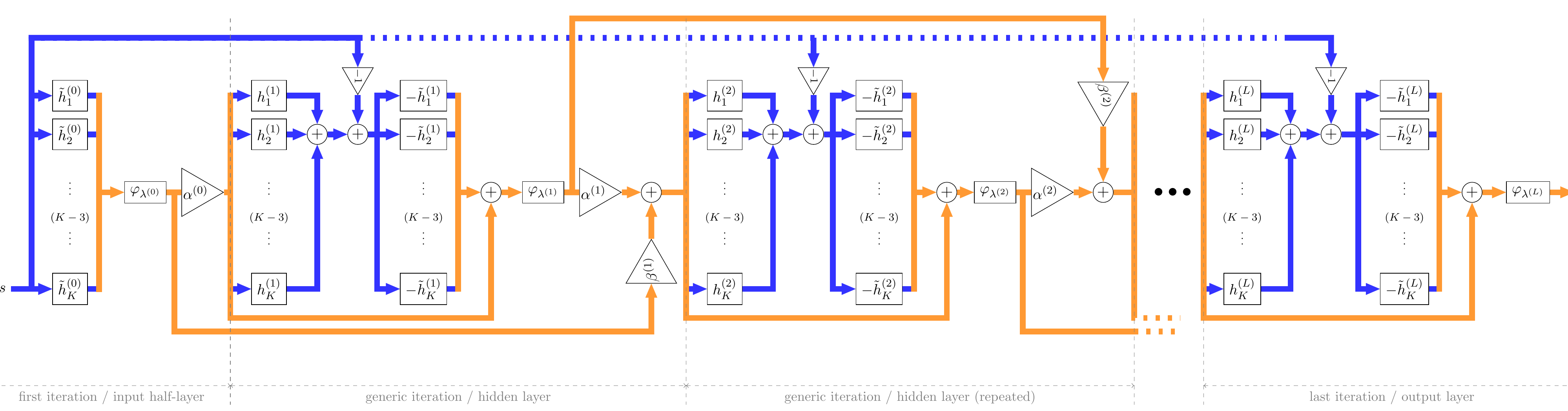}
	
	\vspace{-5pt}
	
	\caption{\label{fig:network} Computation graph corresponding to the accelerated proximal gradient (APG) algorithm to solve convolutional sparse coding (CSC) problems such as \eqref{eq:cc}, when the proximal operator of the regularizer is known in closed form, i.e., $\prox_{\lambda \reg}(\cdot) = \varphi_\lambda(\cdot)$, and $x_1, x_2,\dots,x_K$ are initialized to $0$. Here, $\beta^{(l)} = 1-\alpha^{(l)}$, and $\tilde{h}_k^{(l)}$ is the matched filter to $h_k^{(l)}$. Blue lines represent data flows with size $M\times N$, and orange lines represent data flows with size $M\times N\times K$. When the iterations are not trained, we have that $h_k^{(l)}$ and $\lambda^{(l)}$ do not vary with $l$, and, under technical conditions on the sequence $\alpha^{(l)}$, a cost-function convergence rate of $\mathcal{O}(1/l^2)$ is guaranteed. SpotNet is obtained by training this computation graph, i.e., changing $h_k^{(l)}$, $\alpha^{(l)}$ and $\lambda^{(l)}$, independently for each $l$, to improve the prediction of $x_1,x_2,\dots,x_K$ given an image $s$. }
\end{figure*}

The design of learning-based algorithms to solve optimization problems was pioneered with the learned iterative soft thresholding algorithm (LISTA) proposed in~\cite{Gregor2010}, which was further extended in~\cite{Sprechmann2015,Zhang2018}. The fundamental idea explored in these approaches is to truncate a sparsity-promoting iterative algorithm to a fixed number of parameterized, trainable steps. By learning from a few training samples for which the optimal solution was computed offline, LISTA was shown to speed up the computation of the solution for new samples by orders of magnitude. Recent results suggest that approaches that learn the parameters of a known algorithmic structure establish an optimized tradeoff between convergence speed and reconstruction accuracy in inverse problems~\cite{Giryes2018}, which seems coherent with recent theoretical investigations~\cite{Combettes2018}.  

In addition to optimization problems, learning-based approaches that rely on CNNs are the state of the art in image processing problems~\cite{Krizhevsky2012}. Additionally, CNNs have been gaining traction as the tool of choice for automated medical image analysis~\cite{Litjens2017}. Furthermore, the widespread success of CNNs has motivated the development of optimized hardware and software tools for efficient prototyping, training, and deployment of practical learning-based solutions~\cite{Tensorflow2016}.

In this paper, we use insights from recent results to develop parameterized computation graphs that can be trained offline on a few training pairs of synthetic images and their respective cell locations and particle secretion profiles over time. The trained graphs are then exploited to detect cells' locations and estimate their secretion profiles over time in new synthetic immunoassay images within a fixed computation time. 
On one hand, in Sec.~\ref{sec:spotnet}, we use a finite number of iterations of the algorithm described in~\cite{AguilaPla2017a} as a parameterized graph, which we term~\emph{SpotNet}. 
On the other hand, in Sec.~\ref{sec:cnns}, we use inspiration from the recent successes of CNNs in image analysis to develop a fully convolutional architecture for cell detection, which we refer to as~\emph{ConvNet}. 
In Sec.~\ref{sec:exp}, we provide empirical evidence that the proposed approaches far outperform a human expert in cell detection performance on synthetically generated images. In particular, we obtain an F1-Score exceeding 0.95 for both approaches, compared to a human expert accuracy below $0.75$, in the case of $512\times 512$ pixel images that contain $1250$ cells and are contaminated by Gaussian noise. Furthermore, we provide empirical evidence that, with a similar number of parameters, the complex structure of SpotNet provides a measurable improvement over ConvNet in both the detection and estimation performances.

\section{SpotNet}
\label{sec:spotnet}

\begin{figure*}
    \begin{center}
        \input{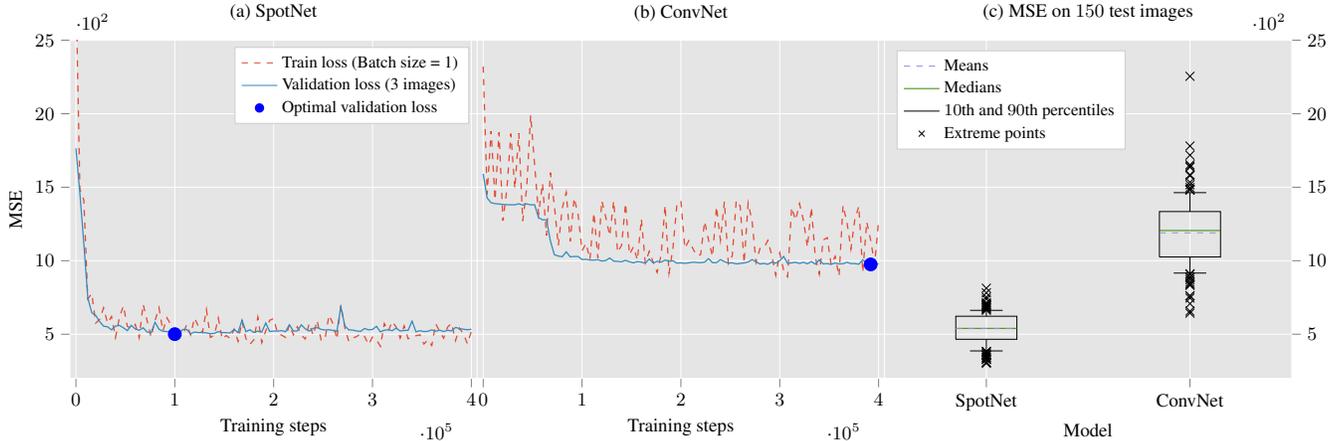}
    \end{center}

    \vspace{-15pt}

\caption{\label{fig:convergence} In (a) and (b), training progress, in terms of training and validation losses on images with $1250$ cells, for the two different models considered. The parameters kept for each model are those corresponding to the optimal validation loss. In (c), statistics for the resulting loss on an independent test database containing $150$ images, composed of groups of $50$ images with $250$, $750$ and $1250$ cells each. As usual, boxes specify the interquartile range (IQR). We observe that 1) SpotNet attains its optimal validation loss four times faster than ConvNet, 2) its test loss is significantly and substantially better, and 3) both in training and test data, SpotNet has a much more stable performance across images, with ConvNet obtaining a test loss with twice the IQR of SpotNet.}
\end{figure*}

A mathematical model for image-based immunoassays was derived in \cite{AguilaPla2017,AguilaPla2018} from the reaction-diffusion-adsorption-desorption process that governs the movement of particles through the assays' medium. In particular, it was shown that, for specific, known, convolutional kernels $\lbrace g_k \rbrace_1^K$, a measured image $s\in\reals_+^{M,N}$ can be expressed as
\begin{equation}
    \label{eq:convolutional_model}
    s \approx \sum_{k=1}^{K} g_k \circledast x_k\,.
\end{equation}
Here, $\circledast$ represents the size-preserving discrete convolution with zero padding. Furthermore, each $x_k\in\reals_+^{M,N}$ is a spatial map of the density of particles secreted from the pixel locations $(m,n)$ during the $k$-th time window in an experiment, i.e., the spatial and temporal particle secretion information one would like to recover. Because the number of cells is much smaller than the number of pixels, each $x_k$ is spatially sparse, and reveals the location of the cell centers. 
Consequently, to recover the $x_k$s from a measured image $s$, the optimization problem \begin{equation} \label{eq:cc}
	\min_{\lbrace x_k \in \reals^{M,N} \rbrace_1^K } \left\lbrace
		\left\| \sum_{k=1}^{K} h_k \circledast x_k -s \right\|_2^2 \!\!\!
		+ \lambda \mathcal{R}(\lbrace x_1, \dots, x_k \rbrace)
	\right\rbrace
\end{equation}
was proposed. This optimization problem fits the convolutional coding model while favouring solutions with structured sparsity through regularization. Specifically, an accelerated proximal gradient (APG) algorithm was proposed to solve \eqref{eq:cc} when the $h_k$s are the known $g_k$s (or an approximation thereof), $\lambda \geq 0$ is a regularization parameter, and $\reg(\cdot)$ is the non-negative group-sparsity regularizer \cite{AguilaPla2017a,AguilaPla2018}.

The APG algorithm to solve \eqref{eq:cc}, described by the computation graph of Fig.~\ref{fig:network}, performs three basic steps at each iteration $l$. First, it performs a  gradient step to minimize the $\ell_2$-norm in \eqref{eq:cc}, i.e., to make the prediction under the convolutional coding model closer to the observed data. Without loss of generality, this gradient step is
\begin{equation*}
    x_k^{(l)} \leftarrow z_k^{(l-1)}  - \tilde{h}_k \circledast \left( \sum_{q=1}^{K} h_q \circledast z^{(l-1)}_q - s \right)\,,
\end{equation*}
where the $z_k^{(l-1)}$s are placeholder variables and each $\tilde{h}_k$ is the matched filter to $h_k$. 
Then, the APG algorithm performs \emph{proximal operator} steps on the $x_k$s, which are non-linear mappings that address the minimization of the regularization term $\lambda \reg(\cdot)$, and are represented by the parameterized non-linear functions $\varphi_\lambda(\cdot)$ in Fig.~\ref{fig:network}. Finally, it performs a Nesterov acceleration step
\begin{equation*}
    z_k^{(l)} \leftarrow x_k^{(l)} + \alpha^{(l)} \left( x_k^{(l)} - x_k^{(l-1)} \right)\,,
\end{equation*}
which updates the placeholder variables $z_k^{(l)}$ for the next iteration.
These three steps are common in convex optimization and, under some conditions on $\alpha^{(l)}$ \cite{Beck2009,Chambolle2015}, guarantee a cost-function convergence rate of $\mathcal{O}(1/l^2)$. Nonetheless, it has been empirically verified that thousands of iterations are required to obtain an accurate estimation of the $x_k$s, which leads to prohibitive computational costs for cell detection and characterization. We refer the interested reader to~\cite{AguilaPla2017a} for a complete mathematical description and empirical evaluation of the APG algorithm to solve \eqref{eq:cc}.

We propose to use the computation graph of a small, fixed number of iterations of the APG algorithm, and to train them to obtain as close an approximation as possible to the particle secretion profiles $\lbrace x_k \rbrace_1^K$. In particular, we propose to take the computation graph in Fig.~\ref{fig:network} for some given $L$ and, for each $l\in\listint{L}$, learn the convolutional kernels $h^{(l)}_k$, the scaling term $\alpha^{(l)}$, and the regularization parameter $\lambda^{(l)}$ so that the loss between the output and the $x_k$s is minimized over a number of training examples. If the learned graph performs and generalizes well, the potential benefits of this approach are immediately clear. First, a fixed number of steps leads to a fixed and known computational complexity. Second, since the convolution kernels are learned, we can further reduce the computational complexity by attempting to learn kernels much smaller than the $g_k$s used by the APG algorithm. Third, the loss function and non-linearity can be chosen arbitrarily as long as they allow gradient-based training of the parameters. In particular, we propose to use 1) a small number of layers, $L=3$, compared to the $10^4$ iterations used in \cite{AguilaPla2017a,AguilaPla2018}, 2) kernels $h_k$ of size $5\times 5$, as compared to the smallest and largest $g_k$s in \cite{AguilaPla2017a,AguilaPla2018}, which were of size $31\times 31$ and $403\times 403$, respectively, and 3) the soft thresholding operator non-linear mapping, and the mean squared error loss function 
\begin{equation}\label{eq:loss}
\mathcal{L}\left(\left\lbrace \left\lbrace h^{(l)}_k \right\rbrace_{k=1}^K, \alpha^{(l)}, \lambda^{(l)}  \right\rbrace_{l=0}^{L}\right) = \frac{1}{K}\sum_{k=1}^K \left\| x^{(L)}_k-x_k\right\|_2^2\,,
\end{equation}
where the $x^{(L)}_k$s are the outputs of the network.

\begin{figure*}
\begin{center}
    \input{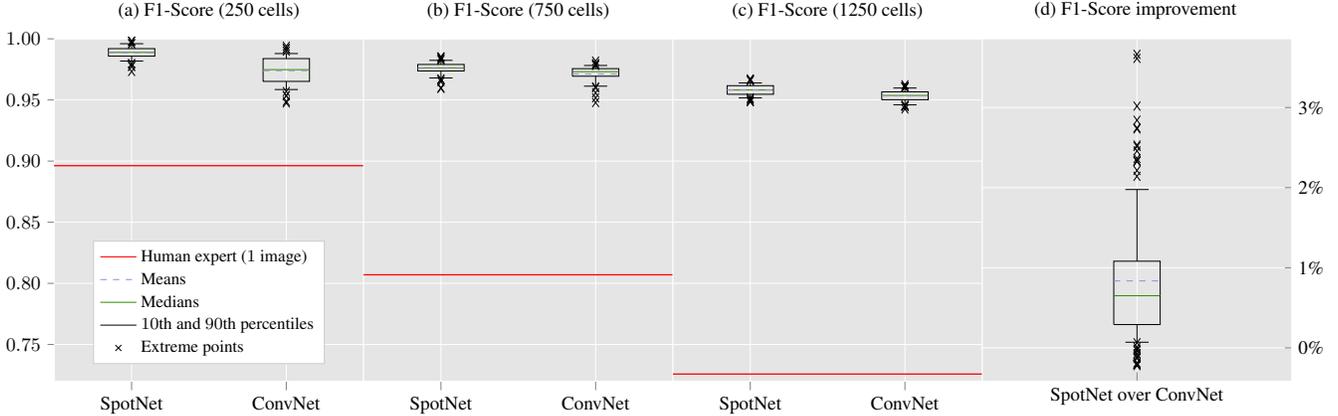}
\end{center}

    \vspace{-15pt}

\caption{\label{fig:detection} In (a), (b) and (c), statistics of the F1-Scores obtained by each model on the $50$ test images for three different scenarios, i.e., (a) $250$, (b) $750$, and (c) $1250$ cells. In (d), statistics of the improvement in F1-Score of SpotNet over ConvNet in the $150$ test images, in percentage points. As usual, boxes specify the IQR. We observe that 1) both SpotNet and ConvNet obtain excellent detection results, well beyond human-expert levels, 2) the distribution of F1-Scores obtained by ConvNet on images with $250$ cells is wider than that on images with $1250$ cells (with $3$ times the IQR of SpotNet in the same category), suggesting overfitting to the latter, and 3) SpotNet yields improved performance over ConvNet in more than $90\%$ of the $150$ test images. }
\end{figure*}

We note that parameterized computation graphs based on proximal gradient algorithms for convolutional sparse coding (CSC) problems have been studied before in \cite{Sreter2017}. The main difference between their work and ours is that 1) our approach is used for recovering the biologically relevant particle secretion profiles for each cell, while \cite{Sreter2017} trains the computation graph on image reconstruction problems by using an additional decoding layer, and 2) we extend the learned CSC studied in \cite{Sreter2017} to include Nesterov acceleration, which is a well-known acceleration technique in convex optimization that introduces skip connections in the computation graph of Fig.~\ref{fig:network}.

\section{CNNs for cell detection}
\label{sec:cnns}

CNNs operate by mimicking the theorized perception within the animal cortex~\cite{Hubel1968}, where the input image is modeled as comprising several locally dependent regions that can further be decomposed into features of varying complexity.
To model these locally dependent regions, a CNN utilizes several trainable~\emph{filters}, $h_k$, $k\in\listint{K}$, that are convolved with an input image $s\in \mathbb{R}_+^{M,N}$ to generate the~\emph{feature maps}
$
    y_k =  \operatorname{\phi}\left( h_k \circledast s + b_k \right)\,,
$
where the $b_k$s are trainable bias variables and $\phi$ is an element-wise differentiable non-linear function.
In effect, each filter encodes a certain learnable feature, and each feature map indicates those regions of the image that contain that particular feature.
By sharing the filters across the image in this manner, CNNs dramatically reduce the amount of parameters required to represent the image compared to fully connected neural networks.
In order to construct more complex features, most CNNs consist of successive layers of filters and their corresponding feature maps arranged in a feed-forward fashion~\cite{Lecun1998}.
A vast number of heuristically designed CNN layouts have been studied in the literature that claim advantages in terms of improving the accuracy of the results, reducing the number of parameters to achieve a target accuracy, improving the robustness to errors in the training data, or a combination of these factors~\cite{Goodfellow2016}.

As discussed in the previous sections, image-based immunoassays generate images comprising several spots of varying shape and instensity.
These spots contain information about the cells, namely, their location and particle secretion profile over time.
Here, we cast the cell detection problem as the task of constructing a CNN, ConvNet, that  extracts these cell-level features by convolving the measured image with a series of trainable filters.
Since the span of each spot is limited to a few tens of pixels, we can use filters that are relatively small (i.e., a few pixels wide) to extract the cell-level features.
It has been shown previously that CNNs composed solely of small convolution filters with a fully connected output layer perform well in the standard image classification benchmarks~\cite{Springenberg2015}.
Therefore, for ConvNet, we exploit the CNN layout of~\cite{Springenberg2015}, but with a modified output scheme that provides an estimate of the spatial maps $x_k$ that contain the cell secretion information.

The ConvNet output depends on the filters and the bias variables, which must be learned to accurately estimate the cell-level features from measured images.
Therefore, to train ConvNet, a few synthetic immunoassay images and their corresponding target values are provided as training data.
Subsequently, the optimal filters and biases are obtained by iteratively updating their values such that the CNN output closely approximates the provided target values.
The ConvNet training phase is typically very computationally intensive and is carried out until some convergence criteria are met. These are usually defined in terms of a loss function that depends on the network output and the target values.
Evaluating the trained ConvNet on input images, however, is relatively inexpensive and requires a fixed computation time, which is suitable for practical cell detection and particle secretion profile estimation.
    
\section{Experimental results}
\label{sec:exp}

	In this section, we provide performance results for cell detection and particle secretion profile estimation using SpotNet and ConvNet on synthetic images. The images are generated following the same procedure as in~\cite{AguilaPla2017a}, and correspond to the noise level $3$ category stated there. Each image has a size of $512\times 512$ pixels, is normalized in the range $[0,255]$, is contaminated by Gaussian noise with $\sigma=4.5\cdot10^{-3}$, and contains $250$, $750$ or $1250$ cells. A detailed description of the simulation steps, along with the complete code to generate our training and test databases, is available in this project's repository \cite{GITHUB}. Our simulator also provides the particle secretion profiles $\lbrace x_k \rbrace_1^K$ used to create each image, where $K=30$.

As stated in Sec.~\ref{sec:spotnet}, we use kernels of size $5\times 5$, a soft thresholding non-linear function and $L=3$ layers for SpotNet. In ConvNet, we use two fully convolutional layers with $6$ and $15$ feature maps each, followed by one convolutional layer that separates each feature map into two, and three per-feature convolutional layers after that. Throughout this structure we use $5\times 5$ filters and one-dimensional biases. In this way, both approaches amount to approximately $210$ convolutions and $7$ scalar parameters, and we obtain a fair comparison in terms of computational cost.
Also in both cases, the loss function is the one specified in \eqref{eq:loss}, i.e., the mean squared error between the network output and the target.

We train SpotNet and ConvNet using seven training images containing $1250$ cells, together with their corresponding particle secretion profiles $\lbrace x_k \rbrace_1^K$. In both cases, we use the Adam optimizer with learning rate $10^{-3}$ and batch size $1$.
At regular intervals, every $4\cdot10^3$ steps, we also calculate the mean loss for three validation images that are not used for training. We stop training after $4\cdot 10^{5}$ steps, and keep the model that has obtained the best validation loss. For more details on, among others, the specific implementation and the training procedure, see this project's repository \cite{GITHUB}.

In Fig.~\ref{fig:convergence}, we observe that SpotNet reaches its best validation loss $4$ times faster than ConvNet, and achieves a validation loss that is half of that obtained by ConvNet. Furthermore, we evaluate the loss of the selected models in a test database of $150$ images composed of groups of $50$ images with $250$, $750$ and $1250$ cells, and observe that the loss value obtained by SpotNet is significantly smaller than that obtained by ConvNet. Finally, SpotNet obtains a more stable performance, obtaining a loss on the test database with half the interquartile range (IQR) of that obtained by ConvNet.
For practical application, this translates into easily trained, robust and accurate estimations of the spatial and temporal particle secretion profiles with SpotNet.

We also quantify the cell detection performance for SpotNet and ConvNet in terms of the F1-score on the three sets of $50$ images in our test database. To do so, we proceed as in \cite{AguilaPla2017a}, i.e., we threshold the local maxima in the temporal mean of the output particle secretion profiles so that the resulting F1-Score is as high as possible in each image. In Fig.~\ref{fig:detection}, we show the statistics of the results, along with the performance of a human expert for a single image.
We observe that both SpotNet and ConvNet perform substantially better than the human expert for all image categories, generalizing well to the image categories that were not present in the training or validation databases. Moreover, SpotNet's performance remains stable across images of the same category, while ConvNet's performance on images with $250$ cells has a higher dispersion (with $3$ times the IQR of SpotNet in the same category), suggesting a slight overfit to the training database. 
Finally, we observe that SpotNet obtains an improved loss with respect to ConvNet in more than $90\%$ of the $150$ test images, reaching improvements of the F1-score of up to $3\%$.

\bibliographystyle{IEEEbib}

{\small
\bibliography{CSCNet}}

\end{document}